# Mott insulators appearing at thickness period corresponding to nesting in $CaRuO_3$

M. Sakoda[1*], H. Nobukane[2], S. Shimoda[3], K. Ichimura[1]

[1]*Department of Applied Physics, Graduate School of Engineering, Hokkaido University, Sapporo 060-8628, Hokkaido, Japan*

[2]*Department of Physics, Graduate School of Science, Hokkaido University, Sapporo 060-0810, Hokkaido, Japan*

[3]*Institute for Catalysis, Hokkaido University, Sapporo 001-0021, Hokkaido, Japan*

Contact author: sakodam@eng.hokudai.ac.jp

In 2021, we discovered a novel size effect with a period of 25 Å on strongly correlated compound $CaRuO_3$. The change in film thickness of only one nanometer leads to an increase in electrical resistivity at 4 K by a factor of several billion. However, the excitation energy of 2.4 eV on insulating $CaRuO_3$ is too large to explain the mechanism by conventional quantum well. In this study, we clarify that the increases in electrical resistivity are accompanied by lattice expansion and caused by the Mott transitions. We measured in-plane X-ray diffraction on $CaRuO_3$ films and found a 24.8 Å periodic thickness oscillation of the lattice spacing $d_{(004)}$. We determined the nesting vector from the Fermi surface and revealed the spin density wave with a period equivalent to the size effect. We concluded that the antiferromagnetic correlation appearing in the boundary conditions triggers the periodic Mott insulators depending on the film thickness.

In a thin film with a thickness on a scale corresponding to the Fermi wavelength, conventional quantum size effects due to quantum wells appear because of the limited dimensions of the electronic system [1-4]. In particular, many compounds have longer Fermi wavelengths than those of mono-elemental metals; therefore, their electronic dimensions can be artificially controlled with great precision. Bulk samples are commonly used to explore novel phenomena involving strongly correlated compounds. Ultrathin films with clear boundaries are valuable for investigating new quantum properties in strongly correlated electron systems. We focused on the perovskite compound $CaRuO_3$ [5-8], which exhibits a strong electronic correlation. We created an artificial low-dimensional electron system by growing ultrathin films with thicknesses on the order of the Fermi wavelengths. The $CaRuO_3$ bulk and thick films are located near the quantum critical point (QCP) [9-11], as indicated by the non-Fermi liquid (NFL) behavior [12-19], large cyclotron effective mass $m_c^* = 4.4\ m_0$ [19], and flat band with W = 30 meV [20]. Although bulk and thick films of $CaRuO_3$ are metallic, ultrathin films of $CaRuO_3$ exhibit an extraordinary size effect that insulates at a thickness period of 25 Å [21]. The change was 9 orders of magnitude larger at low temperatures than the conventional size effect originating from quantum wells [1-4]. In the initial stage of research, even at room temperature, the electrical resistivity could change by as much as 5,000 times, making it a potential size effect device.

This study aims to clarify the mechanism of the extraordinary size effect with a 25 Å period. The temperature dependence of the electrical resistivity of insulated $CaRuO_3$ follows variable range hopping with an activation energy of $E_a = 2.4$ eV [21], which is too large to be explained by the eigenvalue energy of its quantum well. Because the energy is 1-2 orders of magnitude larger than the bandwidth[6] and cannot be understood by the electronic system alone, we assumed that the size effect was accompanied by a change in the crystal structure. The anisotropy of the size effect, which only appears in $CaRuO_3$ (110) growth on the $NdGaO_3$ (110) substrate, should also be considered. $CaRuO_3$ (001) ultrathin films grow on $NdGaO_3$ (001) substrates with sharp reflection high-energy electron diffraction (RHEED) patterns, but no size effect appears [21]. No significant thickness-dependent differences in the crystal structure were qualitatively observed in the Kikuchi patterns obtained by electron backscattered diffraction (EBSD) or RHEED patterns. The details of the structure of the size effect have not been investigated quantitatively. In this study, in-plane X-ray diffraction (XRD) was used to measure the lattice spacing for each film thickness. We clarified the mechanism underlying the remarkable insulating on $CaRuO_3$ in the extraordinary size effect by solving for the differences in each structure.

We grew $CaRuO_3$ thin films with a flat interface using the molecular beam epitaxy (MBE) method (Supplemental Material Fig. S4(b)-(f) [22]). The Laue zones and Kikuchi patterns observed in the RHEED images show a perfectly flat surface and excellent crystallinity, respectively (Fig. 1(a), Supplemental Material Fig. S1, S4(b),(d)-(f) [22]). We measured $\phi$–$2\theta\chi$ scans using in-plane XRD to investigate the crystal structure of the films. The four peaks of (002), (004), (006), and (008), corresponding to the c-plane, were observed for each film thickness (Supplemental Material Fig. S3(a) [22]). A detailed $\phi$–$2\theta\chi$ scan of the most intense (004) peak was obtained. Figure 1(b) shows $\phi$–$2\theta\chi$ scans of $CaRuO_3$ grown at each film thickness. Thickness was measured using the X-ray reflection ratio (XRR) (Supplemental Material Fig. S2 [22]). The (004) peak position was found to depend on the thickness of the film. The blue and orange lines represent the small and large lattice of structure, respectively. The bulk $CaRuO_3$ has a simple rectangular crystal (space group: Pbnm) and a 0.6 % smaller *c*-axis length than the $NdGaO_3$ substrate [23-25] (Supplemental Material Table S1 [22]). Remarkably, the lattice of the $CaRuO_3$ ultrathin films expanded significantly at some film thicknesses. This phenomenon cannot be explained by epitaxial growth, indicating that the lattice of $CaRuO_3$ essentially changes depending on the film thickness. Figure 1(d) shows the lattice spacing $d_{(004)}$ as a function of film thickness. The green dotted line reveals the $d_{(004)} =$ 1.92769(35) Å of $NdGaO_3$ substrate (Supplemental Material Table S1 [22]). A significantly larger $d_{(004)}$ and a smaller $d_{(004)}$ than the substrate were found on $CaRuO_3$. The $d_{(004)}$ of $CaRuO_3$ oscillates with film thickness. A change in film thickness by only a few nanometers leads to expand the lattice by more than 0.3 %. Because conventional $CaRuO_3$ has a shorter *c*-axis than neodymium gallate substrates, the lattice generally decreases in size as it grows epitaxially, approaching the bulk structure (Supplemental Material Fig. S6(a) [22]). Focusing on the minima $d_{(004)}$ in Fig. 1(d), it can be observed that the lattice tends to shrink with increasing film thickness. However, the lattice extensions deviated significantly from the epitaxial rule for some film thicknesses. We numbered the expanded $d_{(004)}$ compared to the substrate, starting from the smallest thickness, and finding the period of T = 24.8 ± 0.4 Å (Supplemental Material Fig. S3(b) [22]). In Fig. 1(c), the periodicity is in accordance with the size effect observed in the electrical resistivity. The thickness accompanied with lattice expansions are consistent with those of insulating $CaRuO_3$ [26, 27, 28]. Comparing the lattice expansion rate with the pressure dependence of the strontium compound $SrRuO_3$ and the layered perovskite $Ca_2RuO_4$, which correspond to negative pressures of -2 GPa [29-31], and -0.4 ~ -0.6 GPa [32-36], respectively. This pressure corresponded to the physical pressure at which the Mott insulator, $Ca_2RuO_4$, was metalized. In contrast, conventional $CaRuO_3$ located near the QCP enters the Mott insulating phase in the Mott-Hubbard model owing to the negative pressure. Consequently, it was confirmed that the large activation energy observed in the extraordinary size effect originated from the Mott gap.

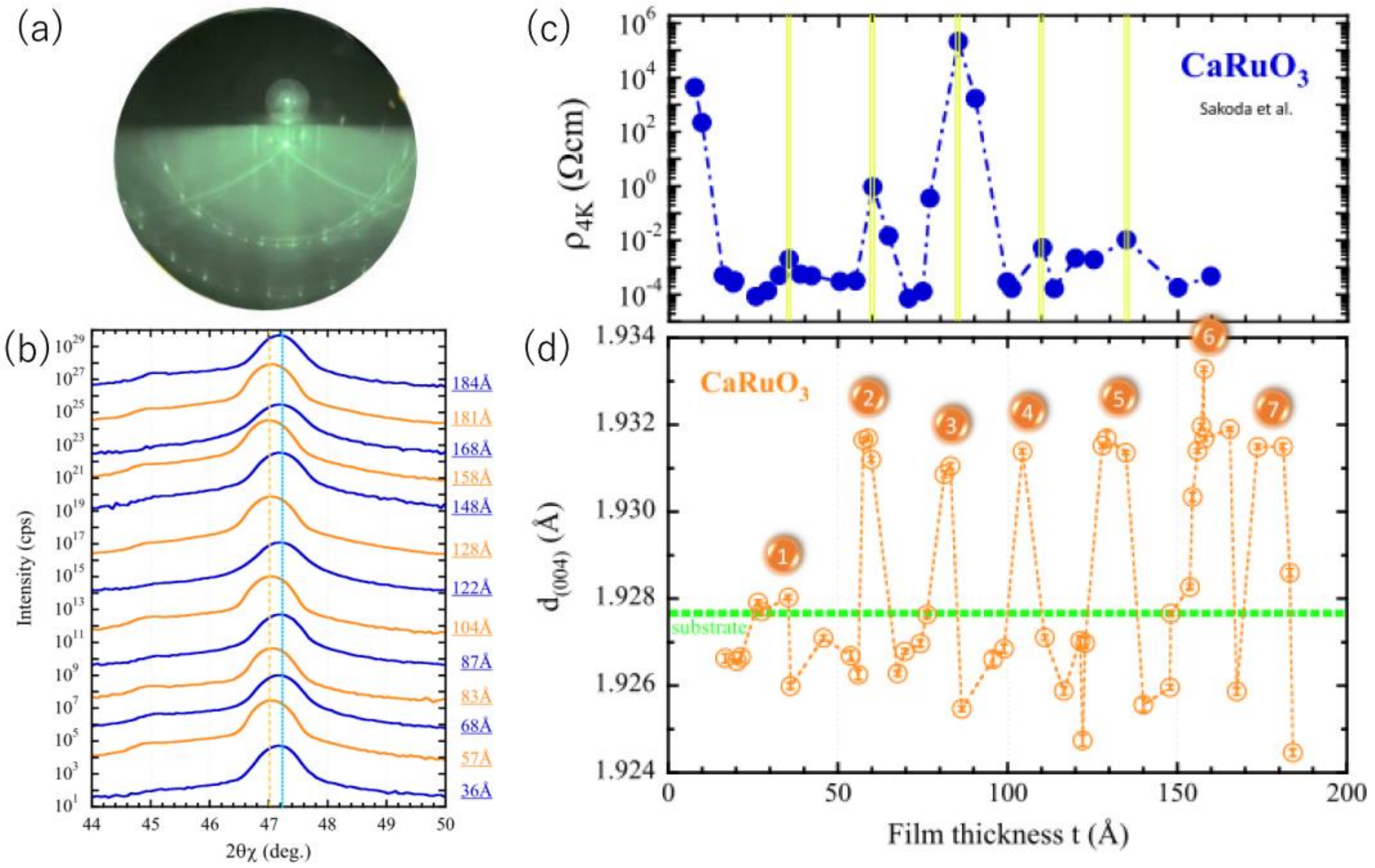

**FIG. 1.** Thickness dependence of crystal structure of $CaRuO_3$. (a) RHEED pattern of $CaRuO_3$ ultrathin film. (b) in-plane XRD of various thicknesses of $CaRuO_3$. They are shifted up every 100 times for visibility, starting from the smallest film thickness. (c) Electrical resistivity[5] at 4 K $\rho_{4K}$ and (d) lattice spacing $d_{(004)}$ as a function of film thickness of $CaRuO_3$. The yellow lines show a maximum of $\rho_{4K}$. The green dotted line indicates $d_{(004)}$ of substrate. The peaks are numbered in order from thinner.

The structure of the $TO_6$ (T = transition metals) octahedron in perovskite compounds dominates physical properties [37-41]. We observe the $RuO_6$ structure in the expanded $CaRuO_3$ with $d_{(004)}$ = 1.932 Å by using the scanning transmission electron microscope (STEM). Figures 2(a),(b), and (c) show schematic crystal structures of $CaRuO_3$, STEM-annular bright field (ABF), and high-angle annular dark field (HAADF) images near the interface with the $NdGaO_3$ substrate, respectively. The crystal period at the interface was clear, and there were no dislocations, indicating excellent epitaxial properties (Supplemental Material Fig. S5(a),(b) [22]). Fig. 2(d) shows the tilt angle $\theta$ of $RuO_6$ (transverse axis), which is expressed as the distance per $RuO_6$ cell from the interface (longitudinal axis). The $\theta$ is smaller than $\theta = 11.2$ of the bulk $CaRuO_3$. A small $\theta$ value indicated that the expanded $CaRuO_3$ lattice was under compressive strain in the growth plane (Supplemental Material Fig. S6(b) [22]). In contrast, the reported tilt angle of $CaRuO_3$ in a conventional lattice is as large as $\theta > 20°$ and is under strong tensile strain [42] (Supplemental Material Fig. S6(a) [22]). In Fig. 2(e), the expansion of the lattice forced $RuO_6$ to stand upright in the growth direction. The intercept of 6.9 ± 1.6 Å for the period shown in Supplemental Material Fig. S3(b) [22] indicates that the two $RuO_6$ octahedrons (= $CaRuO_3$ unit cell) directly above the substrate, which are significantly affected by epitaxial strain, do not contribute to the size effect.

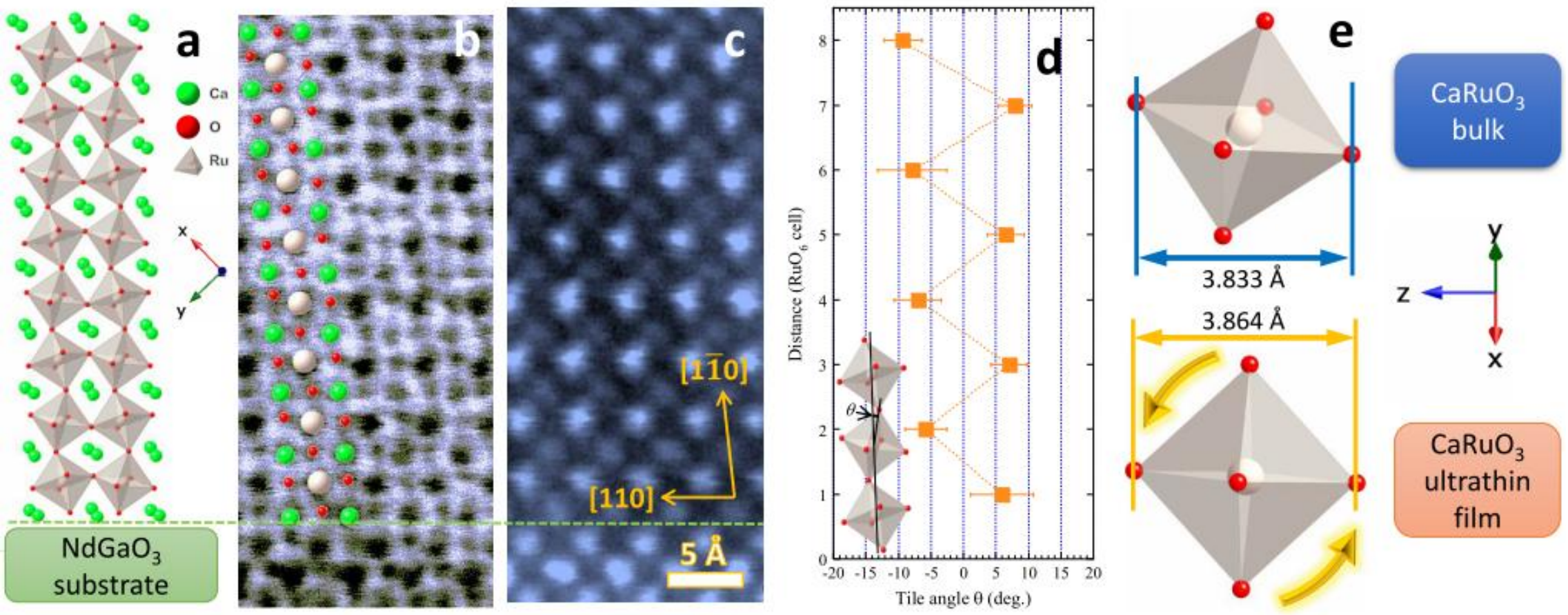

**FIG. 2.** Structure of expanded $CaRuO_3$. (a) Schematic of the crystal structure of $CaRuO_3$. Corresponding (b) STEM-ABF, and (c) STEM-HAADF images at the interface of $NdGaO_3$ substrate. The green line shows the interface between $CaRuO_3$ and $NdGaO_3$ substrate. (d) Plots of the tilt angle of $RuO_6$ octahedral structure. The inset schematic shows the angle $\theta$ in connected $RuO_6$ octahedra. (e) Comparison of tilt angle and lattice length on $CaRuO_3$ bulk and expanded $CaRuO_3$ ultrathin film.

Magnetoresistance (MR) and Hall effects were measured to clarify the differences in magnetic response and electronic structure between conventional $CaRuO_3$ with a thickness of 54 Å and lattice-expanded $CaRuO_3$ with a thickness of 58 Å. Figure 3(a) shows the MR with a temperature $T = 2$ K and current $I$ along the $c$-axis (Supplemental Material Fig. S7 [22]). Conventional $CaRuO_3$ exhibits positive MR in both the magnetic field directions perpendicular and parallel to the film surface, which is due to scattering based on the typical Fermi surface (FS) effect [43]. In contrast, the expanded $CaRuO_3$ exhibited a negative MR in all magnetic field directions; this is the typical behavior of Mott insulators in suppressing the antiferromagnetic state using a magnetic field. Fig. 3(b) shows a fitting with $(\mu_0 H)^2$ for $\mu_0 H < 5$ T. The $\Delta\rho/\rho_0$ is proportional to $-(\mu_0 H)^n$ ($n = 1.92$ for H ⊥ surface, 2.19 for H // surface, 2.00 for H // c), which follow the Mott hopping diagram fits very well [44, 45]. Figure 3(c) shows the angle dependence of MR at $T = 2$ K and $\mu_0 H = 14$ T. The $a$ and $b$ are the easy and difficult axes, respectively, as determined from the crystal orientation confirmed by EBSD, as shown in Fig. 3(d). The antiferromagnetic spins of the Mott insulator were aligned along the $a$-axis.

Figure 3(e) shows the Hall resistivity as a function of the magnetic field at 2 K. The Hall resistivity of conventional $CaRuO_3$ is small, with electron-like carriers caused by the large multi-connected FS occupied by electrons [17,20]. Although the expanded $CaRuO_3$ also had electron-like carriers, its Hall resistivity decreased significantly. This indicates that the number of carriers decreases with lattice expansion. In Fig. 3(e) inset, the reduction in the Hall coefficient with increasing magnetic field indicates that Mott antiferromagnetism is suppressed, which brings carriers to the conventional metallic state. Fig. 3(f) shows the magnetoresistance of $CaRuO_3$ film with thickness of t = 35 Å. We found a jump at magnetic fields as small as H ~ 100 Oe. The change in scattering is considered to be due to the suppression of the antiferromagnetic moment. The suppression of scattering at small magnetic fields is supposed to be attributed to spin density waves. The rapidity of the change suggests that the original ferromagnetism of $CaRuO_3$ affects the change in antiferromagnetic moment [46, 47].

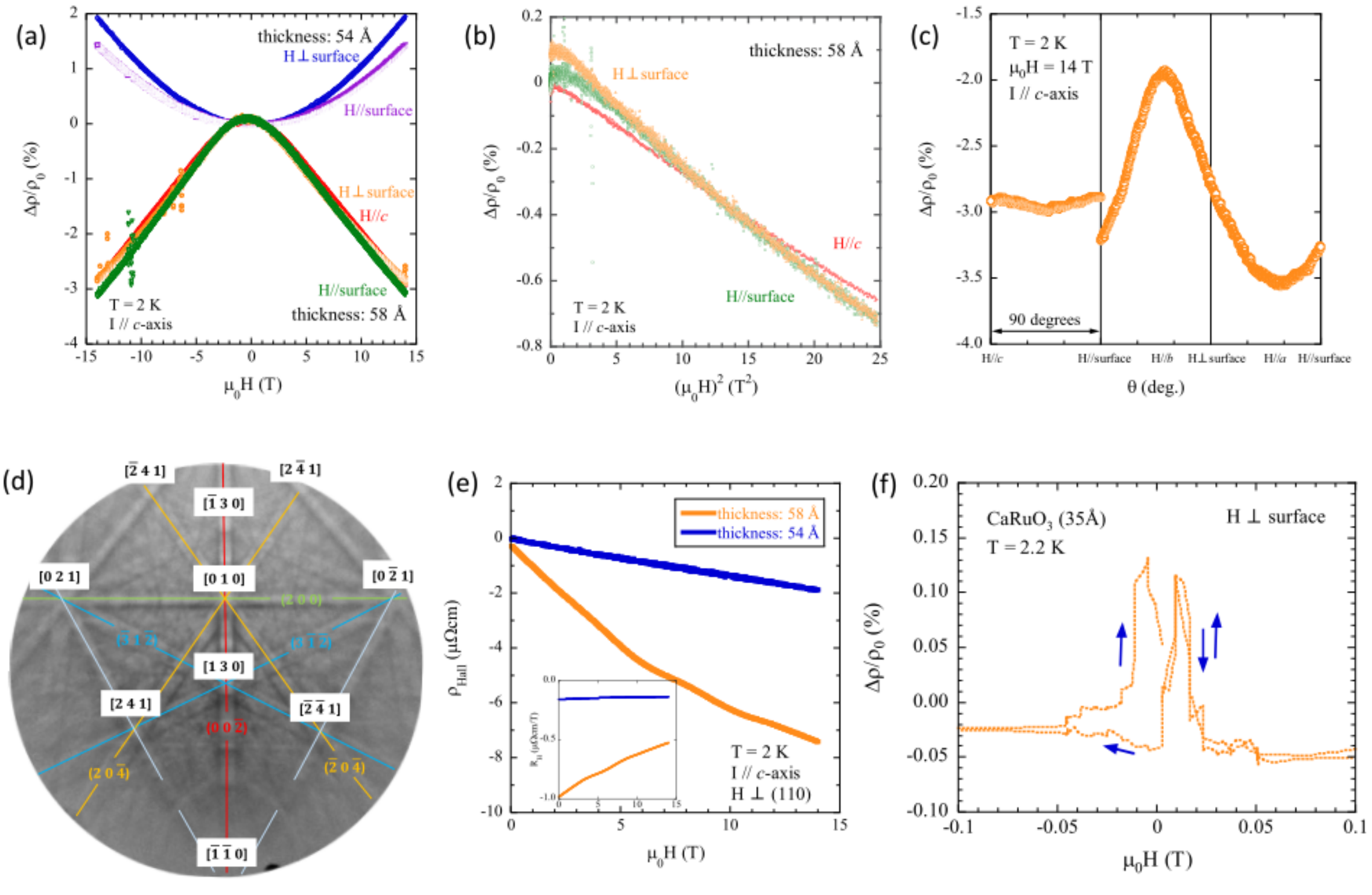

**FIG. 3.** Properties in magnet field. (a) MR of conventional $CaRuO_3$ with a thickness of 54 Å and expanded $CaRuO_3$ with a thickness of 58 Å. (c) MR as a function of angle on expanded $CaRuO_3$. (d) Assigned crystal direction by EBSD pattern on $CaRuO_3$. (e) Hall resistivity as a function of magnetic flux density. The inset shows their Hall coefficients.

We clarified the origin of the 25 Å period in the size effect. The length was not consistent with the unit lattice $d_{(110)}$ = 7.689 Å in the growth direction of $CaRuO_3$, implying the appearance of a new period in the crystal. The FSs of $CaRuO_3$ were reported using angle-resolved photoemission spectroscopy (ARPES) [20]. We found a nesting vector $\boldsymbol{Q} = h\boldsymbol{a}^* + k\boldsymbol{b}^*$ ($h = k = 0.28 \sim 0.33$) in a square pole FS open to $c^*$ axis (Fig. 4). The energy gain of the electronic system is large because the dominant FS disappears, accompanied by the opening of the bandgap in many states. Indeed, the decrease in carriers associated with lattice expansion, as indicated by the Hall resistivity (Fig. 3(e)), indicates the disappearance of the FS. In real space, it corresponds to a density wave with length $\ell$ = 23.5 ~ 27.6 Å period to the growth direction [110] of the $CaRuO_3$ films. The period is incommensurate to the $CaRuO_3$ unit cell since it is longer than three cells (= 23.1 Å) and shorter than four cells (= 30.8 Å). The length exactly matches the 25 Å period of the size effect in Fig. 1(c),(d). Standing waves can form in nanoscale ultrathin films only at film thicknesses that satisfy the boundary condition of multiples of 25 Å. Nesting does not occur in a film of thickness that produces a fraction of the density wavelength because the energy gap cannot be opened across the film. Even if the film thickness part corresponding to a multiple of 25 Å receives gain, the remainder of the fraction does not. Such boundary conditions have not been considered in large-scale bulk samples. This quantum phenomenon could occur only in ultrathin films, where the fraction is relatively large.

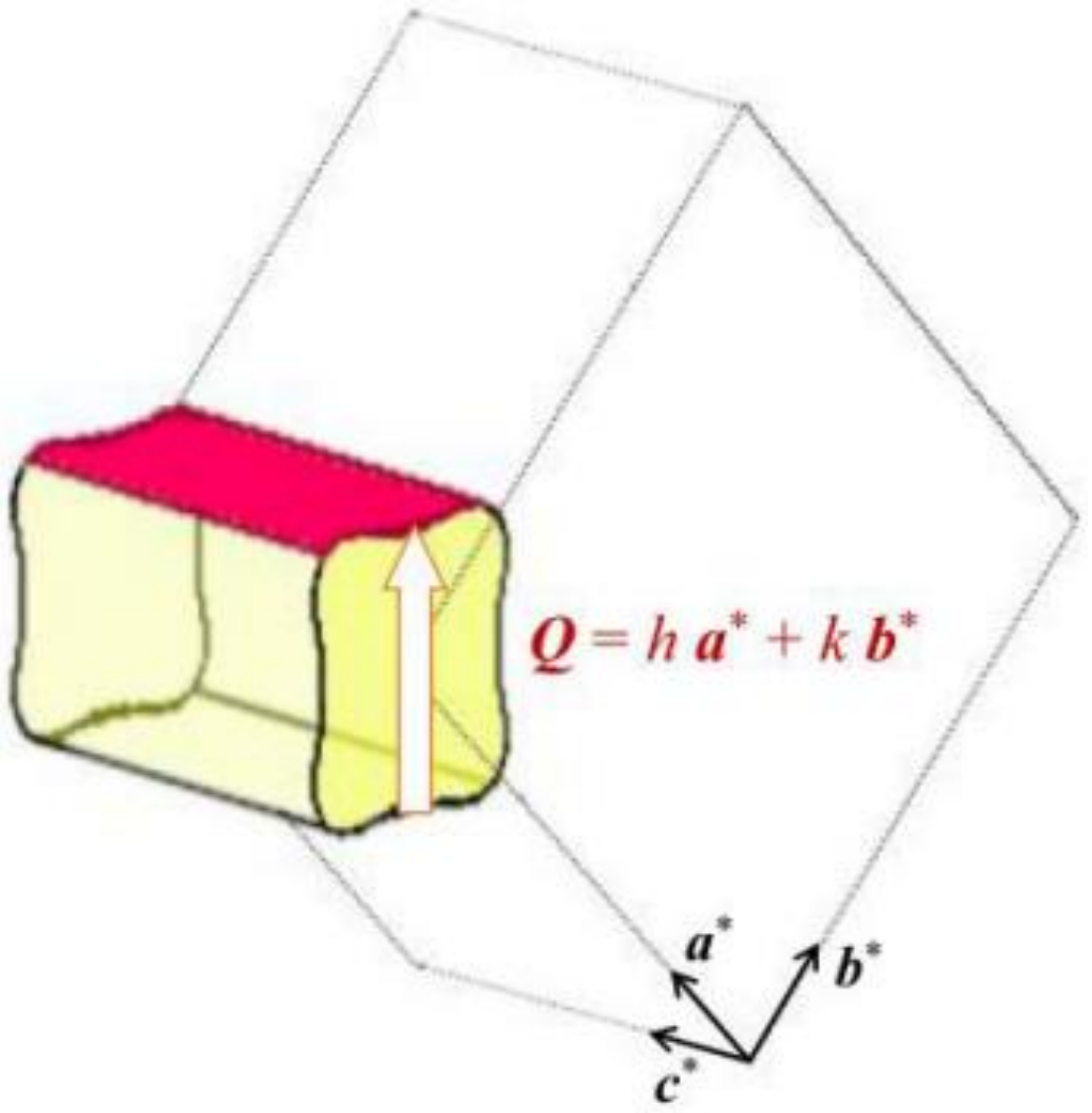

**FIG. 4.** Nesting vector of $CaRuO_3$. The dotted lines indicate the 1st Brillouin zone. The square pole FS, is shown at its corner. The arrow indicates the nesting vector $\boldsymbol{Q} = h\boldsymbol{a}^* + k\boldsymbol{b}^*$.

The TEM diffraction and STEM-fast Fourier transform (FFT) images of the expanded $CaRuO_3$ showed only lattice spots (Supplemental Material Fig. S8 [22]). No satellite spots were caused by the appearance of new electrostatic potentials. This result rules out the possibility of charge order, such as CDW. Therefore, it can be concluded that density waves created by spins (SDW) are the origin of nesting. Since the energy gain of the electron system associated with nesting is on the order of δK ~ 0.01 eV, an even stronger electron correlation must be considered to explain the activation energy of $E_a$ = 2.4 eV in the extraordinary size effect. In such strongly correlated electron systems, where an FS that can make SDW and another FS that does not make nesting exist simultaneously, the magnetic moment accompanied by the SDW assists in the formation of Mott insulators through exchange interactions [48]. $CaRuO_3$ also has a square-pole FS that produces SDW and another FS with a flat band that does not produce nesting. A heavy FS constitutes a strongly correlated electron system that can be localized in the Mott phase or itinerant. Because conventional $CaRuO_3$ is weakly Ferromagnetism [17,49-53], it is incompatible with Mott insulators that form antiferromagnetic spins. Here, the appearance of antiferromagnetic correlations caused by the SDW triggers the construction of the Mott insulator. It is suggested that the decrease in the density of states at the Fermi energy $D(E_F)$ associated with the disappearance of the FS also contributes to the lattice expansion by suppressing the shielding effect and enhancing the Coulomb repulsion at the ruthenium sites. In fact, no size effect was observed for [001] growth because of no nesting properties [21], this is a unique size effect that appears only for the [110] growth, where a nesting vector occurs.

Figure 5 shows a schematic of the size effect that periodically transforms a Mott insulator depending on the film thickness. Fig. 5(a),(b) show the STEM-HAADF images and the corresponding crystal structure, respectively. Fig. 5(c)-(h) show the electronic structures on ruthenium site for each film thickness. Each oscillation indicates the formation of a standing wave of SDW in a limited thickness toward the growth direction. The SDW can be considered to be one-dimensional in the plane-perpendicular direction. In $CaRuO_3$ with thicknesses corresponding to multiples of the period $T$ = 25 Å (~ $RuO_6$ x7 and x14) shown in Fig. 5(d),(g), the standing wave of the SDW triggers the Mott transitions in which the lattice is expanded and carriers are localized. As indicated by the angular dependence of the magnetoresistance (Fig. 3(c)), the antiferromagnetic moment is oriented along $a$-axis. As density waves cannot form in a film with a thickness that disagrees with the SDW period, the film enters a conventional itinerant state without lattice expansion (Fig. 5(c),(e),(f),(h)). In Fig. 5(i) [9], (j), $CaRuO_3$ shows an interesting quantum movement to come and go between the original quantum critical region and the Mott-insulating phase as a function of film thickness. Quantum-critical properties, which have been studied using pressure and magnetic fields, can be controlled using a new concept of scale.

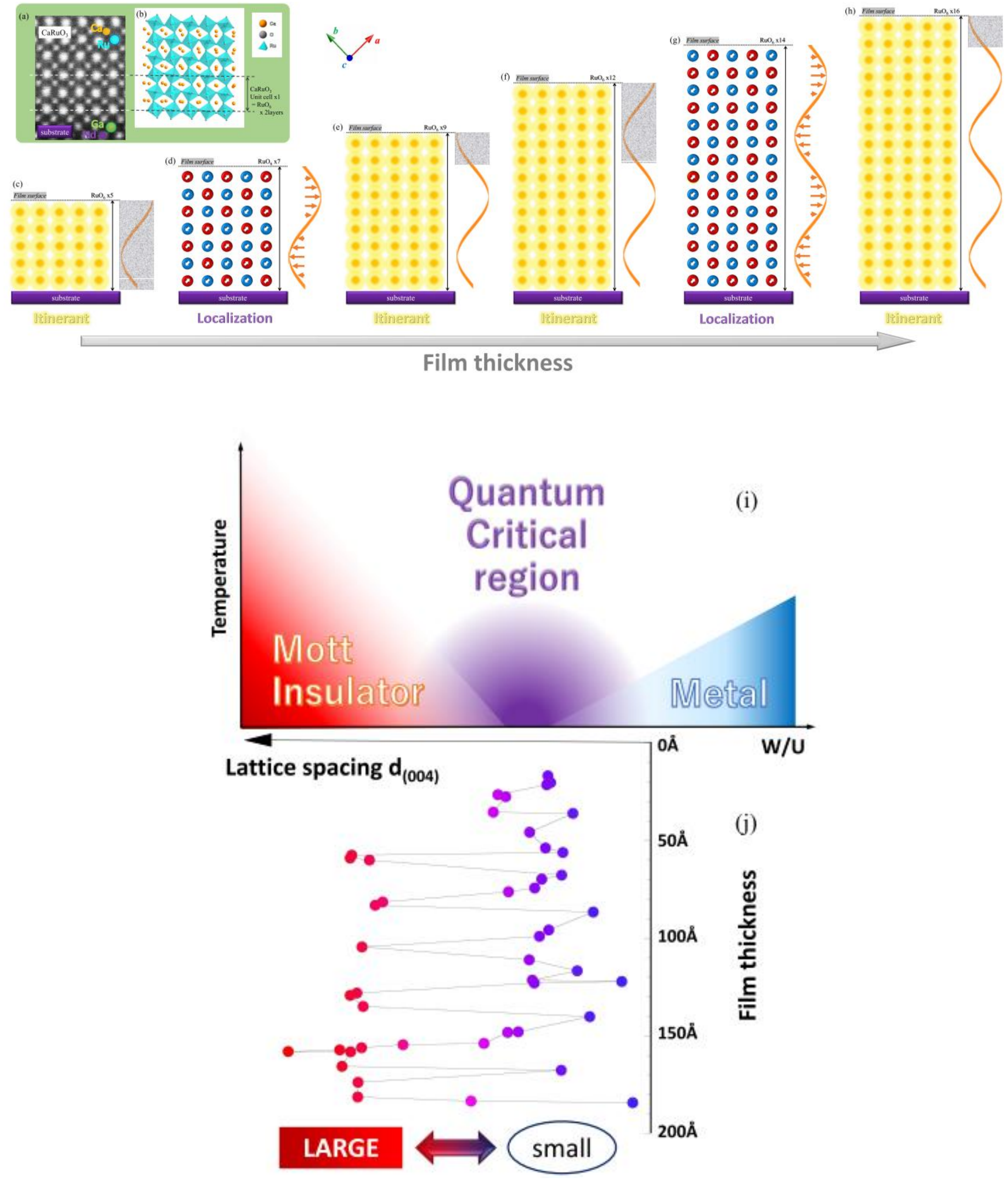

**FIG. 5.** Origin of extraordinary size effect. (a) STEM-HAADF, and (b) schematic structure of $CaRuO_3$. They are viewed in the c-plane, with the vertical indicating the direction of growth. (c)-(h) state of $RuO_6$ site in $CaRuO_3$ ultrathin films for various thicknesses. (d) and (g) electrons are localized to be Mott insulators with anti-ferro magnetic along the *a*-axis. The others show an itinerant state to have conductivity. (i) in the Mott-Hubbard model, longitudinal and transverse axes represent temperature and bandwidth $W$ over Coulomb repulsion $U$, respectively. The bulk $CaRuO_3$ is located in the quantum critical region [9]. (j) lattice spacing $d_{(004)}$ in Fig. 1(d) corresponds to the $W/U$ of the Mott-Hubbard model. The longitudinal axis indicates the film

thickness. $CaRuO_3$ lattice expands at 25 Å period and transitions to Mott insulator.

In summary, we have prepared $CaRuO_3$ ultrathin films with clear boundaries using the MBE method. It is found that the lattice oscillates with a period of T = 24.8 ± 0.4 Å depending on the thickness of the films using in-plane XRD. The lattice expansions are linked to an increase in electrical resistivity, indicating the Mott insulation of $CaRuO_3$ due to negative pressure. The expansion of the lattice due to the upright $RuO_6$ octahedrons was observed from STEM observations. The MR of the expanded $CaRuO_3$ is negative, and the easy axis is $a$. The Hall effect indicates that the number of carriers decreases with the lattice expansion. A nesting vector causing SDW with period $\ell$ = 23.5 ~ 27.6 Å to the growth direction [110] was found. The SDW antiferromagnetic correlation triggers the appearance of the Mott transition at a thickness period of 25 Å. This strong electron correlation results in an increase in electrical resistivity by several billion times.

*Acknowledgements*-The authors thank Professors S. Tanda, M. Naito, S. Kashimoto, K. Yamazaki, and Y. Sasaki for valuable discussion and support. This study was supported by the Joint Usage/Research Center for Catalysis. A part of this work was conducted at the Laboratory of XPS Analysis (Mr. K. Suzuki), Laboratory of Nano-Micro Materials Analysis (Mr. A. Sawa), High-Voltage Electron Microscope Laboratory (Mr. T. Tanioka), and Joint-use Facilities, Hokkaido University. This work was partially supported by the "Advanced Research Infrastructure for Materials and Nanotechnology in Japan (ARIM)" of the Ministry of Education, Culture, Sports, Science, and Technology (MEXT), Proposal Number JPMXP1224HK0067. This study was financially supported by the TEPCO Memorial Foundation, Ito Science Foundation, Telecom Advanced Technology Research Support Center, and Nippon Sheet Glass Foundation for Materials Science and Engineering.

**Supplemental information for**

# Mott insulators appearing at a thickness period corresponding to nesting in $CaRuO_3$

M. Sakoda[1*], H. Nobukane[2], S. Shimoda[3], K. Ichimura[1]

[1]*Department of Applied Physics, Graduate School of Engineering, Hokkaido University, Sapporo 060-8628, Hokkaido, Japan*

[2]*Department of Physics, Graduate School of Science, Hokkaido University, Sapporo 060-0810, Hokkaido, Japan*

[3]*Institute for Catalysis, Hokkaido University, Sapporo 001-0021, Hokkaido, Japan*

Contact author: sakodam@eng.hokudai.ac.jp

## Methods

### Growth of films

$CaRuO_3$ films were grown using a home-built MBE system with a basal pressure of $\approx 1 \times 10^{-5}$ Pa. The growth temperature is 800 °C. Ruthenium (Ru) was evaporated with an electron beam using Hydra e-guns (Thermionics Laboratory, Inc.). Calcium (Ca) was evaporated by electron beam heating or resistive heating using the Knudsen cell (Vecco, Inc.) with dual filaments with a typical temperature of T.C. = 530 °C with constant power mode. Typical emission currents were 100 and 15 mA for Ru and Ca, respectively, with an acceleration voltage of 90 kV. The Ru molecular beam was controlled by electron impact emission spectroscopy (EIES) using the Guardian controller (INFICON, Co., Ltd.). The Ca molecular beam was controlled and monitored using the EIES for electron and resistive heating. The EIES sensors were calibrated using a quartz crystal microbalance with a Cygnus 2 deposition controller (INFICON, Co., Ltd.) inserted at the substrate position. $CaRuO_3$ thin films were grown at a rate of 25 Å/min. The target rate of the Ru molecular beam was 0.1 Å/sec. The target rate of the calcium molecular beam was 0.45 Å/s for the supplied molecular ratio of Ca/Ru = 1.4. The surface of $CaRuO_3$ films grows the flattest under deposition conditions with a supply ratio of Ca/Ru = 1.4 [42]. The $CaRuO_3$ films had a stoichiometric ratio of Ca:Ru = 1:1, as determined by Energy Dispersive X-ray Spectroscopy (EDS). Excess calcium re-evaporates from the substrates owing to its high vapor pressure.

The oxides were grown by supplying ozone $O_3$ using a conventional ozonizer at a concentration of ≈ 14 %. The partial pressure of oxygen $p_{O2}$ was measured using a quadrupole mass electrometer with RGA300 (Stanford Research Systems, Inc.). The optimized growth condition of oxygen partial pressure was $p_{O2} \approx 4 \times 10^{-4}$ Torr for all $CaRuO_3$ films [54]. The optimized growth temperature was 800 °C [54]. The details of the homemade MBE system are described elsewhere [55-57].

The substrate was neodymium gallate $NdGaO_3$ with a (110) surface supplied by Shinko-sha Co., Ltd. The typical size is about 3 mm × 5 mm. The average values of substrate supplied by Shinkosha are $d_{(004)}$ = 1927.69 ± 0.35 mÅ, $d_{(008)}$ = 963.56 ± 0.05 mÅ, and full width at half maximum FWHM = 0.3965 ± 0.0110 degrees.

The surface roughness of the $NdGaO_3$ substrate was ensured to be below 2 Å. $NdGaO_3$ substrates were annealed on the holder in an MBE chamber at 800 °C for 1 h prior to deposition. $CaRuO_3$ with a (110) surface was grown on a substrate of $NdGaO_3$ with a (110) surface [54,58]. Because the crystal structure of $CaRuO_3$ is very similar to that of the $NdGaO_3$ substrate, the $CaRuO_3$ films grew epitaxially along the crystal orientation on the substrates. Accordingly, $CaRuO_3$ autoregulated compounds to stoichiometric ratios on the $NdGaO_3$ substrate. Highly crystalline flat $CaRuO_3$ ultrathin films were reproducibly grown using homemade MBE equipment.

The surfaces of the films were observed in situ via RHEED using RDA-003G (R-DEC Co., Ltd.). The acceleration voltage was set at 19 kV. The typical emission current was 35-45 μA. The RHEED pattern in Fig. 1 shows the first to third Laue zones of the semicircular row of points and streak patterns. These images indicate that the thin films contained a 'perfect flat surface' with no defects. The Kikuchi bands and lines were also observed, indicating excellent crystallinity. It was possible to obtain sharp RHEED patterns with good reproducibility for all film thicknesses [21,42,54].

**Measurements**

Initially, we attempted to measure the $CaRuO_3$ lattice spacing using out-of-plane XRD. But, the lattice $d_{(110)}$ = 7.722 Å of the $NdGaO_3$ substrate and the lattice $d_{(110)}$ = 7.689 Å of the $CaRuO_3$ were too close to each other. Therefore, the $CaRuO_3$ peak was not measured accurately because it was hidden by the $NdGaO_3$ peak or appeared to be small at the foot of the $NdGaO_3$ peak. Consequently, we used in-plane XRD, which has a shallow X-ray incidence angle and can precisely detect only nano-order thickness on the sample surface. The lattice spacing of the $CaRuO_3$ (001) surface was measured by $\phi$–$2\theta\chi$ scans of the in-plane XRD using SmartLab (Rigaku Co., Ltd.). The incidence angle was optimized to $\omega$ = 0.2-0.4 degrees for each sample to obtain the highest intensity. The standard deviation was 0.38 mÅ, and the standard error was 0.19 mÅ when the same $NdGaO_3$ substrate was measured four times. The error produced by each measurement was sufficiently small compared with the scale of the size effect. Since the Kβ filter is not used, the peak derived from the Kβ line also appears to be on the low angle side of the main peak due to the Kα line (Fig. S3(a)). The FWHM of all $CaRuO_3$ was the same as that of the substrate, indicating excellent crystallinity (Fig. S3(c)).

The thicknesses of the $CaRuO_3$ films were measured by XRR using SmartLab (Rigaku Co., Ltd.). A clear oscillation was observed in the raw data. (Fig. S2(a)). The thicknesses of a film measured by XRR and STEM were 291.49 Å and 288.1 Å, respectively (Fig. S2(b),(c)). The difference is as small as 3.4 Å (1.2 %).

EBSD was also performed to determine the crystal direction of the $CaRuO_3$ film using field-emission scanning electron microscopy (SEM) JSM-7000F (JEOL Co., Ltd.).

Atomic force microscopy (AFM) (SPA-400, Hitachi High Technologies Co., Ltd.) was used to measure the surface roughness of the $CaRuO_3$ films. The surface structure of the $CaRuO_3$ thin film was transformed after the film thicknesses of 200 Å and 400 Å (Fig. S4(b)). Since the surface roughness is small at film thicknesses below 150 Å, $CaRuO_3$ grows along the substrate surface. In the AFM image with a film thickness of 165 Å, the terrace structure is becoming faintly visible (Fig. S4(g)). The terrace structure was clearly observed in the AFM image above 200A film thickness, causing steps within a single thin film (Fig. S4(h)). It is considered that there is a partial mixture of conventional and expanded $CaRuO_3$ within the same thin film, which resembles the current path reported for the size effect of electrical resistivity. At a film thickness of about 100 Å, the insulation is suppressed to make small current paths in the film (fig1(c)). Conversely, since XRD detects larger area structures more preferentially, the size effect clearly appears up to a film thickness of nearly 200 Å (Fig. 1(d)). Above 400 Å film thickness, the average surface roughness is $R_a$ = 8.0 ~ 9.4 Å, and the surface flatness is significantly reduced. The streak splits in the RHEED pattern indicate steps in the film (Fig. S4(a)). The Kikuchi band in the RHEED pattern indicates good crystallinity.

**TEM observations**

The samples were quarried from the films and thinned by focused ion beam (FIB) milling with gallium ions using a JIB-4601F Multi Beam System (JEOL Co., Ltd.). After the $CaRuO_3$ thin film was prepared, gold was deposited 100 nanometers thick to protect the surface from etching by gallium ions. Furthermore, the samples were thinned via a voltage of 200–1,000 V argon-ion gentle milling using an IV-8 (Technoorg Linda, Inc.). The argon-ion milling is also effective in removing residual gallium ions from the surface.

STEM experiments were conducted using an electron microscope (JEM-ARM 200F; JEOL Co., Ltd.) at room temperature. The instrument was equipped with double spherical aberration correctors and a cold-field emission gun. The microscope was operated at an acceleration voltage of 200 kV for all the experiments. The STEM images were collected with a 14–17 mrad convergent angle with a 20 μm condenser aperture. The collection angles were 12–24 and 68–280 mrad for ABF and HAADF images. The atoms are visibly sharp in the HAADF image because the scattering angle is large, and the effect of impurities is minimal. However, because the scattering of oxygen, a lighter element, was small, the ABF images, which were observed at a lower angle, were used in conjunction. To determine the oxygen positions that constitute the $RuO_6$ octahedron, we observed the *c* direction, where the oxygen appears to overlap in the same position. Although some images were also captured with the *c*-axis horizontal, information on $RuO_6$ was scarce because the two oxygen atoms appeared to be linked together.

The TEM diffraction was performed at room temperature on an electron microscope (JEM-ARM 200F; JEOL Co., Ltd.). The acceleration voltages were in the range of 50-200 kV. The camera length was 100 cm for all the experiments. Nano-beam electron diffraction (NBD) and selected-area electron diffraction (SAED) with a 100-nanometer diameter aperture were used to observe the $CaRuO_3$ ultrathin films. Fig. S8(c) was obtained using the NBD. The electron beam was narrowed down to a range of ~10 nm using the NBD and observed at a position directly below the gold, where no gold scattering was observed. When the aperture diameter of the SAED exceeded the thickness of $CaRuO_3$, the substrate was also observed simultaneously. The diffraction spots of $CaRuO_3$ and the NGO substrate were so similar that they did not interrupt the satellite spots.

**Low-temperature properties**

We measured the MR and Hall effect at temperatures as low as 2 K with a magnetic field of up to 14 T using a physical property measurement system (PPMS, Quantum Design, Inc.). The measurement setup is shown in Fig. S7. The films were covered with a metal mask and milled with argon ions to pattern the six terminals; this enables us to set the Hall terminals perpendicular to the source-drain current. Less noise owing to the source-drain voltage allows the Hall voltage to be measured accurately. We soldered silver wires to the 5 probes with indium to measure the MR and Hall effects. Some neodymium gallate substrates contain slight magnetic impurities that cause the substrate position to shift under high magnetic fields. The thin film was fixed by wrapping a Kapton tape around the base.

In the B//[110] direction, where the Hall effect was measured, the orbit in the multi-connected FS contributed to the Hall effect [20]. In this direction, the open-cylinder FS contributed slightly to the Hall effect. Therefore, it is reasonable to estimate the carrier density *n* by one-carrier model $R_H = -1/ne$. Conventional $CaRuO_3$ is metallic with $n = 4.01 \times 10^{21}$ electrons/cm$^3$. In contrast, the carrier density of the expanded $CaRuO_3$ decreased to $n = 6.37 \times 10^{20}$ electrons/cm$^3$.

**Method references**

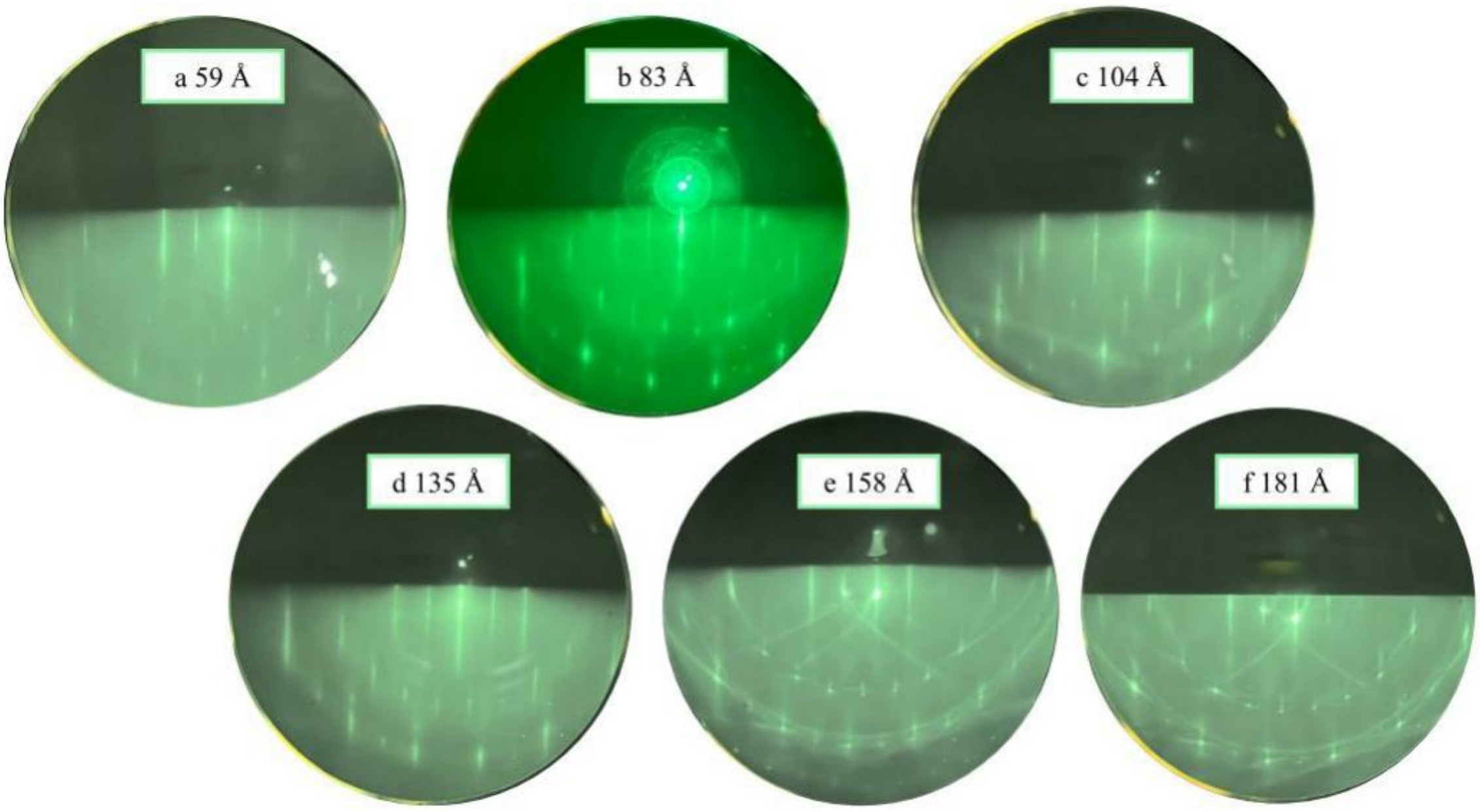

**Fig. S1.** The RHEED patterns in the expanded CRO films. The Laue zone and the Kikuchi patterns are shown.

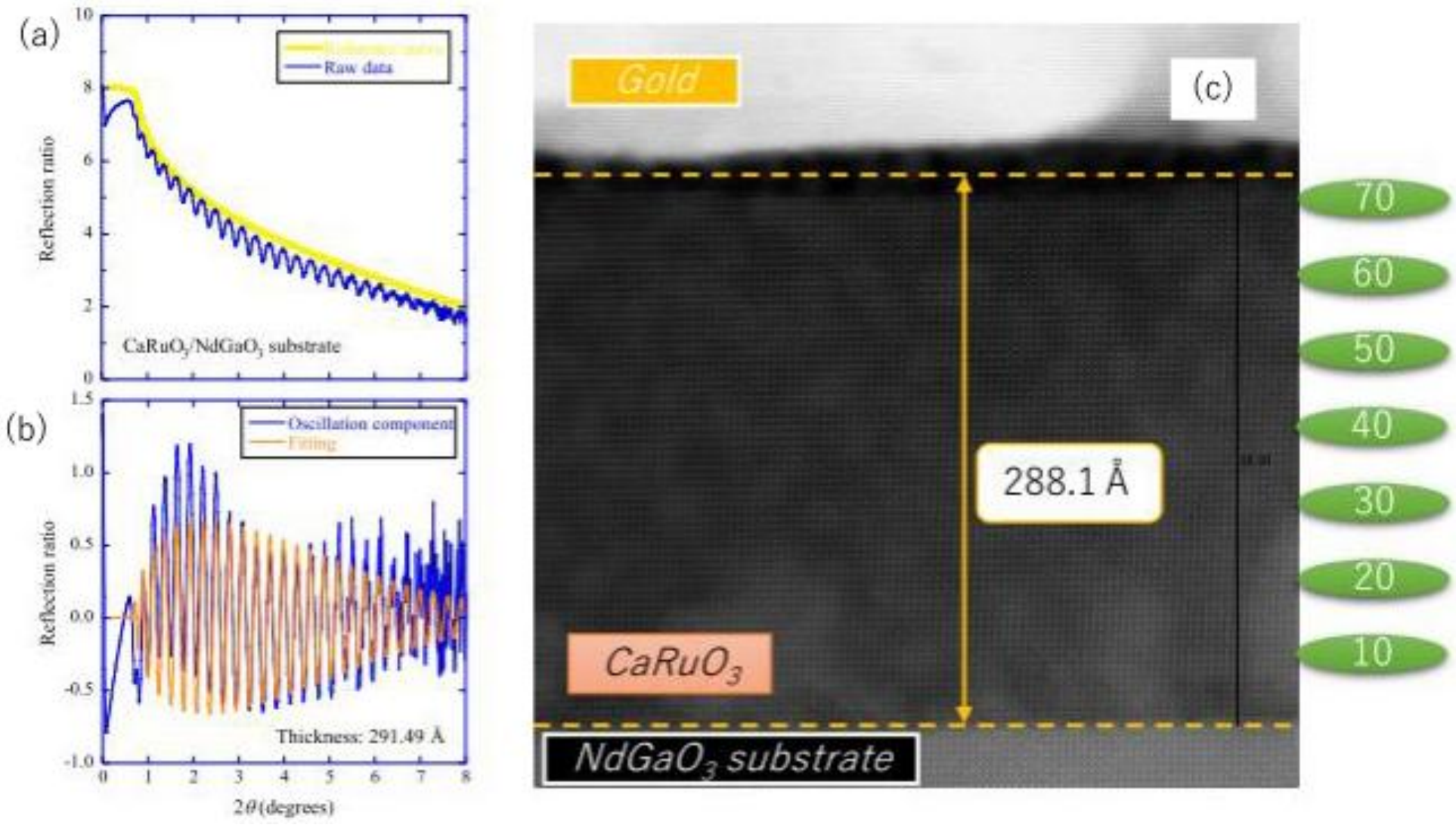

**Fig. S2.** (a) Angular dependence of XRR. The blue and yellow lines represent the raw data and reference curves, respectively. (b) Vibrational component in blue and its fitting in orange are shown. (c) Corresponding film thickness observed in the STEM-HAADF image. $CaRuO_3$ is sandwiched between the $NdGaO_3$ substrate and gold protective layer. The steps of Ru atoms from the interface are numbered on the right.

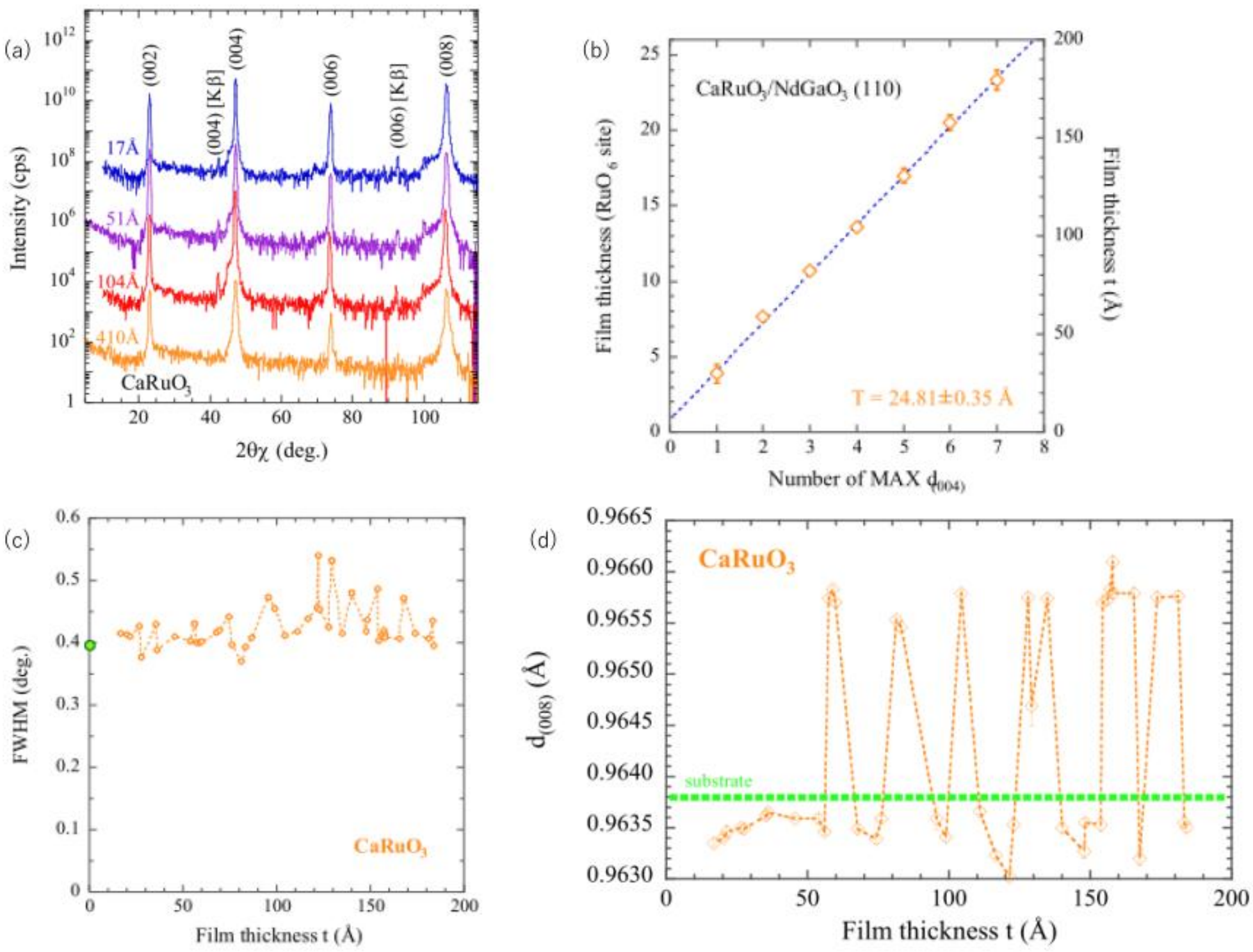

**Fig. 3.** (a) ϕ–2θχ scans measured by in-plane-XRD on $CaRuO_3$ films of several thicknesses. (b) Plots of average film thickness on MAX $d_{(004)}$. (c) FWHM of the (004) peak plotted as a function of $CaRuO_3$ film thickness. (d) $d_{(008)}$ as a function of film thickness of $CaRuO_3$. The green dotted line indicates $d_{(008)}$ of the substrate.

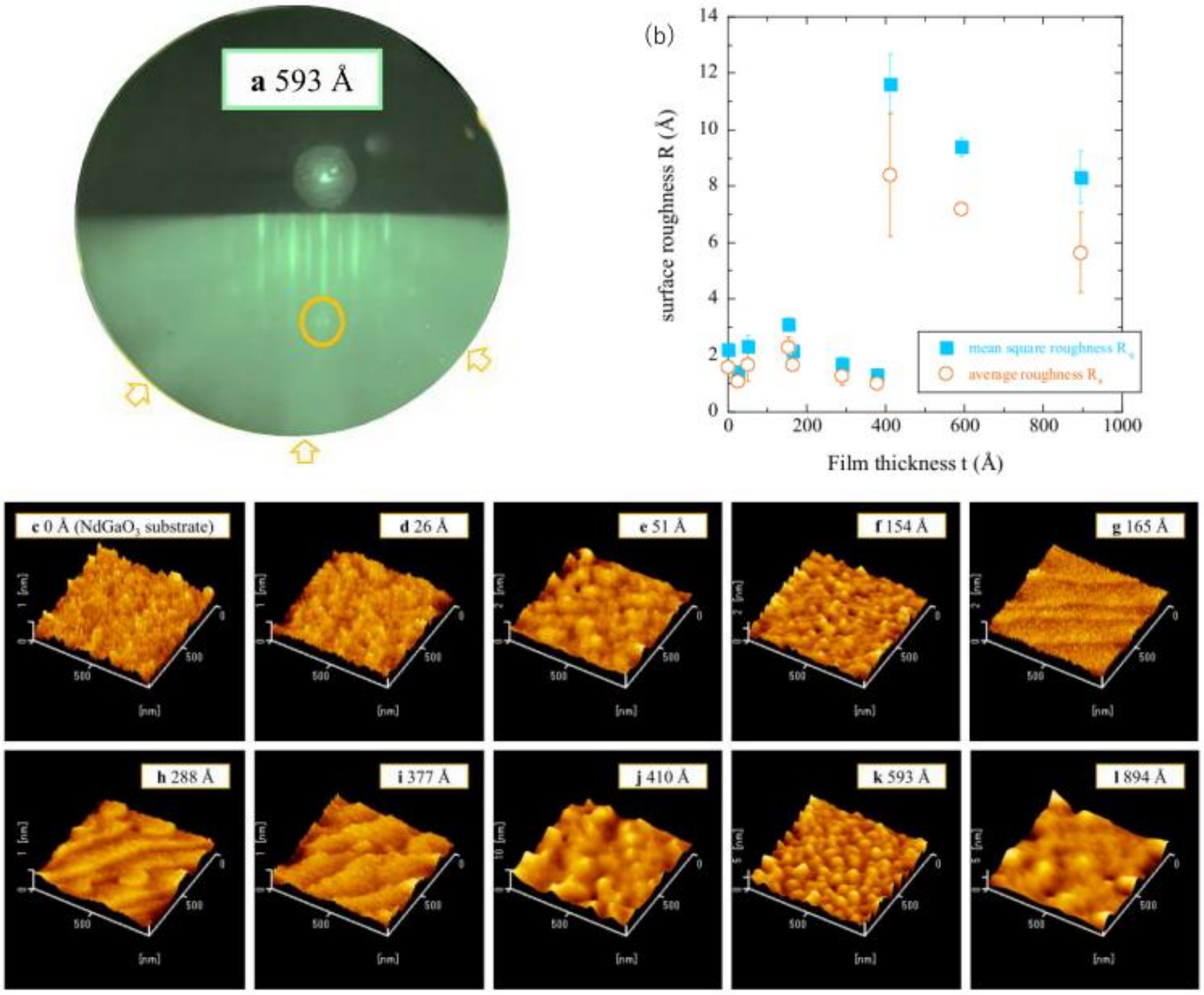

**Fig. S4.** (a) RHEED pattern of $CaRuO_3$ with a thickness of 593 Å. The yellow circle and arrows indicate the splitting streak and Kikuchi patterns, respectively. (b) Thickness dependence of average roughness $R_a$ and mean square roughness $R_q$. The surfaces of (c) $NdGaO_3$ substrate, and (d)-(l) $CaRuO_3$ films with various thicknesses observed by AFM.

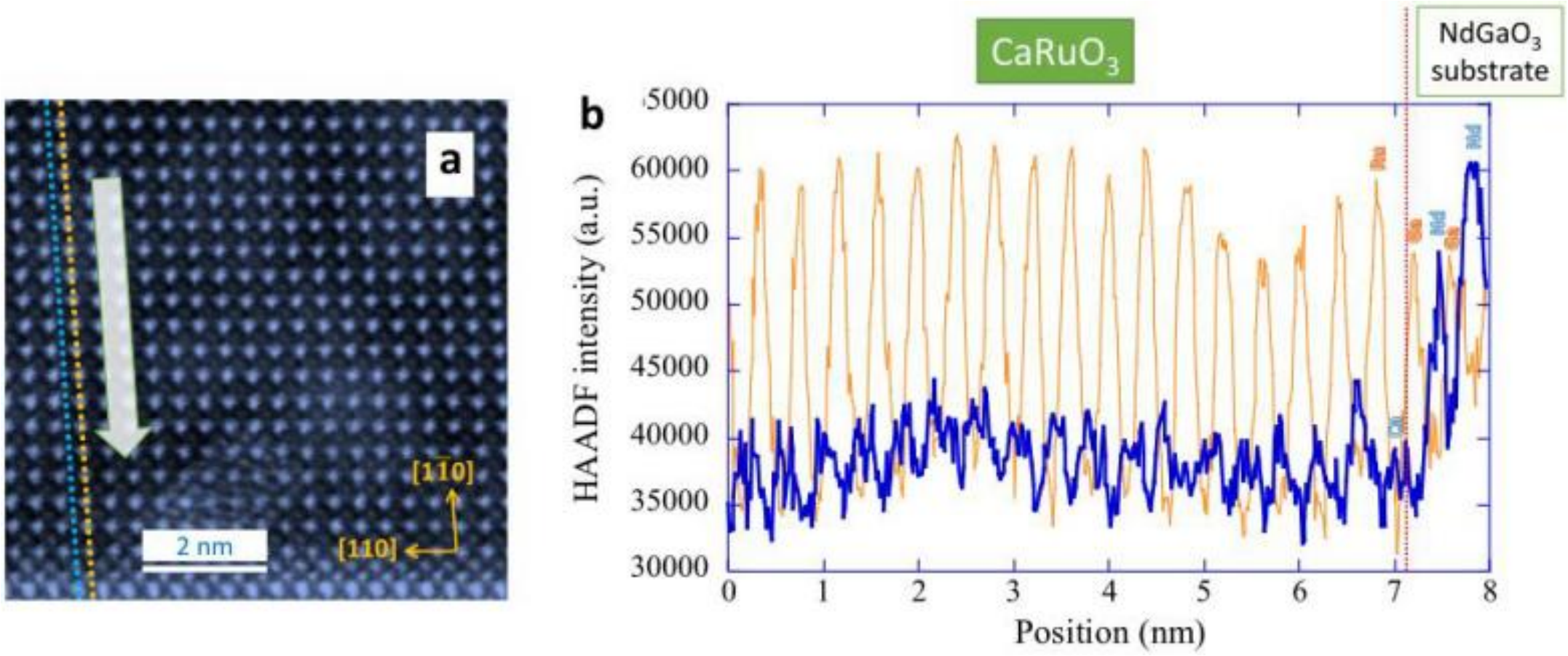

**Fig. S5.** (a) HAADF-STEM image of $CaRuO_3$ taken near the $NdGaO_3$ substrates interfaces. (b) HAADF-intensity line profiles of $CaRuO_3/NdGaO_3$ at positions indicated by vertical yellow and blue dash lines in Fig. S5(a). Peaks corresponding to Nd, Ga, Ca, and Ru-site position were indexed for each component.

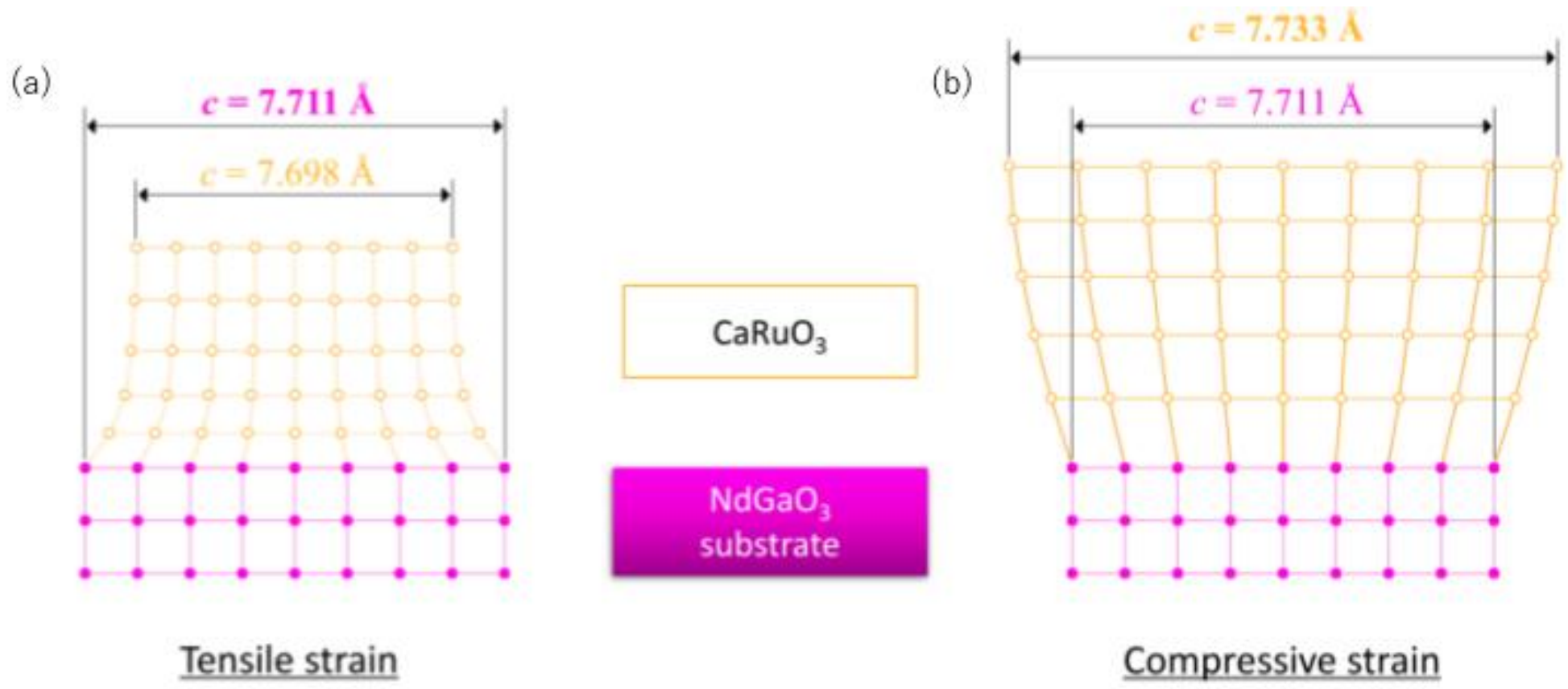

**Fig. S6.** (a) Tensile and (b) compressive strain for conventional and expanded $CaRuO_3$ epitaxial growth on $NdGaO_3$ substrates, respectively.

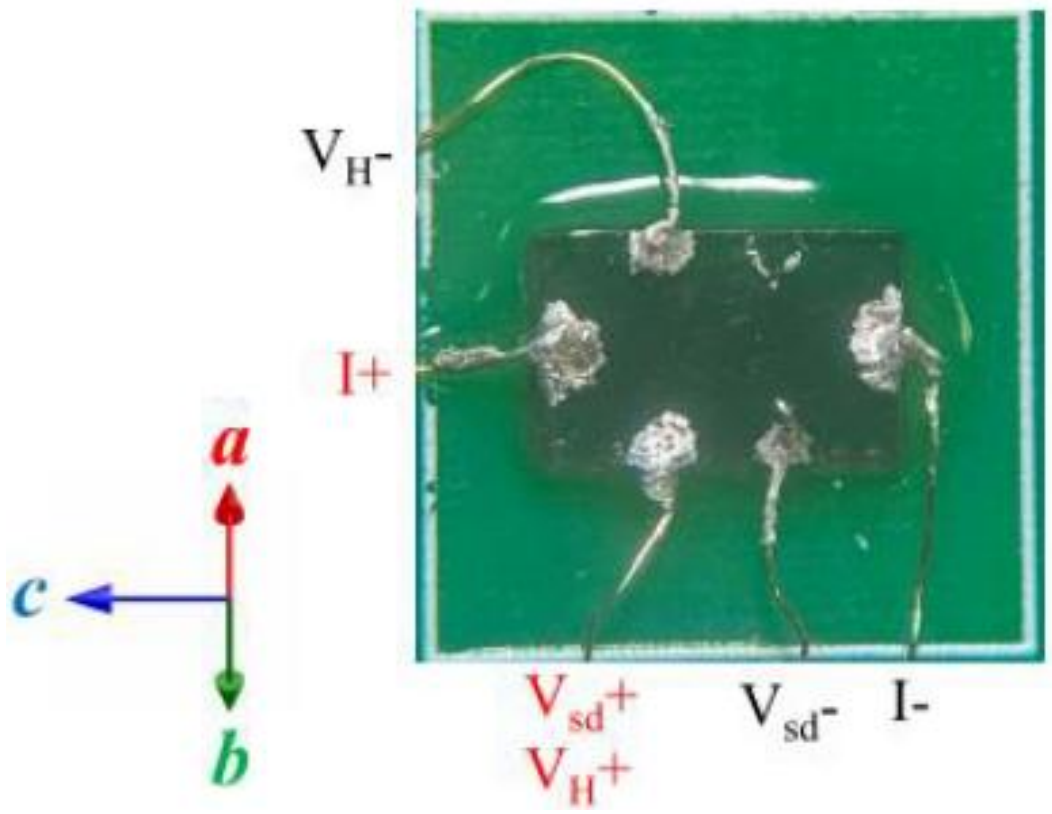

**Fig. S7.** Configuration of $CaRuO_3$ film on MR and Hall effects measurements. The five terminals bonded to the film are the current I, source-drain voltage $V_{sd}$, and Hall voltage $V_H$. Current direction towards the c-axis. The Hall voltage is measured along the [110] direction.

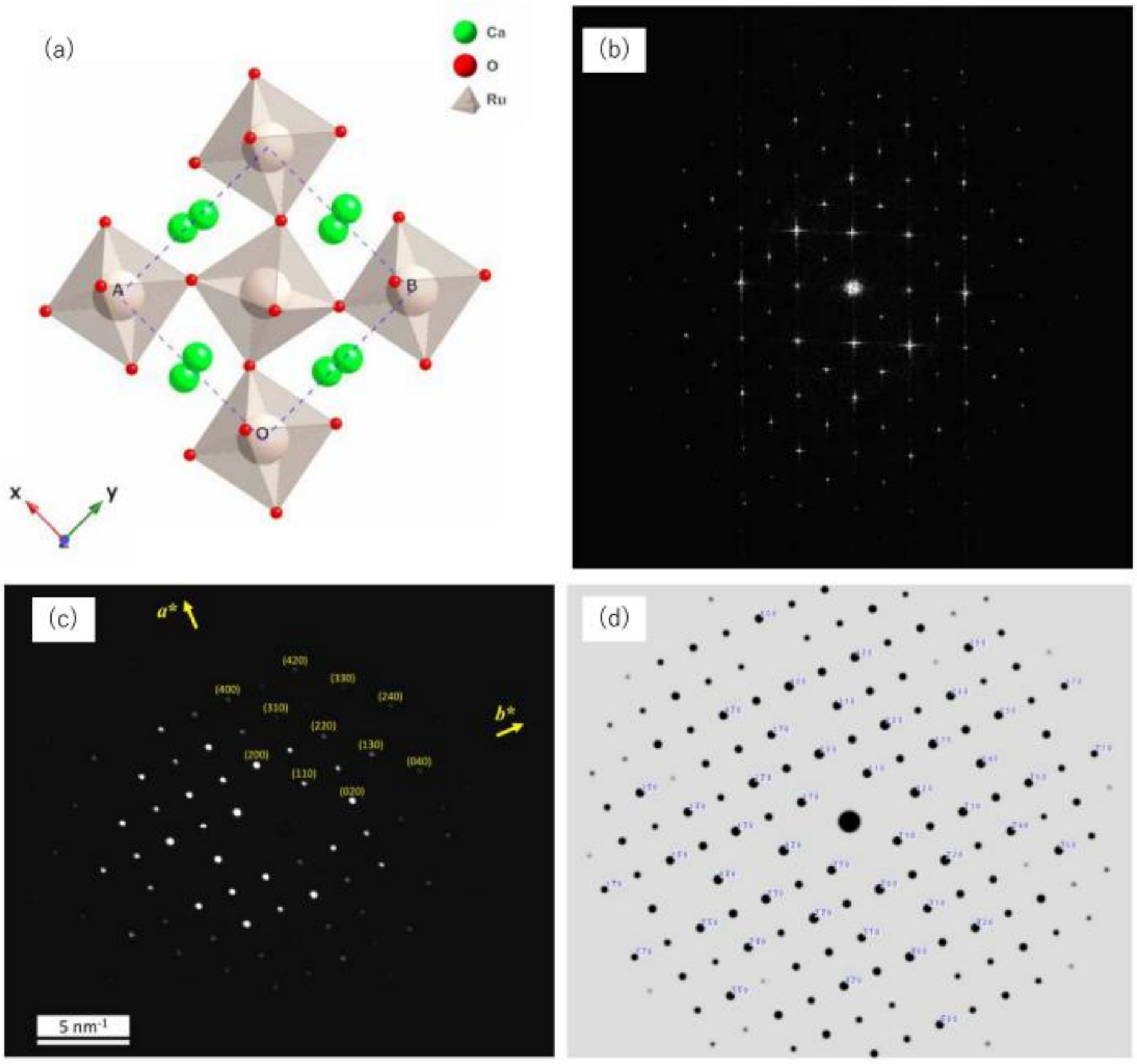

**Fig. S8.** (a) Crystal structure of $CaRuO_3$ observed in the c-plane. The blue dotted lines represent the unit cells. The green spheres, red spheres, and white octahedrons represent calcium, oxygen, and ruthenium, respectively. (b) Corresponding FFT image calculated from the STEM-ABF image. Each point represents the diffraction of the unit lattice. The bright spots are due to the $RuO_6$ periodicity, which is included two in the unit length. (c) TEM diffraction image observed in the $c^*$ plane. The $a^*$ and $b^*$ directions are also shown in the figure. (d) Corresponding simulated TEM diffraction image for reference.

**Table S1.** Summary of lattice spacing $d_{(004)}$ and $d_{(008)}$. Bulk $CaRuO_3$ lattices were reported by five groups. Bulk $NdGaO_3$ lattices were reported by two groups. $d_{(004)}$ and $d_{(008)}$ were calculated from the lattice constant in the literature. We measured the lattices of two $NdGaO_3$ substrates supplied by Shinkosha Co. Ltd. We measured one of the badges multiple times to check the reproducibility. For reference, we measured the $NdGaO_3$ lattice supplied by K&R Ltd..

| [redacted] | | | |
|---|---|---|---|
| Bulk $CaRuO_3$ | 1.918 | 0.959 | [5] |
| | 1.912 | 0.956 | [6] |
| | 1.915 | 0.958 | [7] |
| | 1.918 | 0.959 | [8] |
| Bulk $NdGaO_3$ | 1.927 | 0.953 | [23] |
| | 1.927 | 0.964 | [24] |
| $NdGaO_3$ substrate | 1.92792±0.00010 | 0.96359±0.00005 | This studying #1 (*supplied by Shinkosha) |
| | 1.92771±0.00008 | 0.96349±0.00003 | This studying #2 (*supplied by Shinkosha) |
| | 1.92805±0.00008 | 0.96363±0.00003 | |
| | 1.92761±0.00008 | 0.96356±0.00003 | |
| | 1.92714±0.00010 | 0.96354±0.00004 | |
| | 1.92669±0.00008 | 0.96346±0.00003 | This studying #3 (*supplied by K&R) |